\documentclass{article}

\usepackage{spconf,amsmath,graphicx}
\usepackage{subfigure}
\usepackage{latexsym}
\usepackage{amssymb}
\usepackage{amsfonts}
\usepackage{epsfig}
\usepackage{cite}
\usepackage{calc}
\usepackage{color}
\usepackage{theorem}
\usepackage{url}
\usepackage{subfigure}
\usepackage{cases}
\usepackage{verbatim}
\usepackage{array}
\usepackage{dcolumn}
\usepackage{multirow}
\usepackage{tabularx}
\usepackage{setspace}
\usepackage{algorithmicx} 
\usepackage{algorithm}
\usepackage{caption}
\usepackage{algpseudocode}
\usepackage{lipsum}
\usepackage{scrextend}
\addtokomafont{labelinglabel}{\sffamily}

\usepackage[T1]{fontenc}
\usepackage{graphicx}
\usepackage{threeparttable}
\usepackage{epsf,epsfig}
\usepackage{amsmath}
\usepackage{siunitx}
\usepackage{amssymb}
\usepackage{dsfont}
\usepackage{subfigure}
\usepackage{color}
\usepackage{algorithm}
\usepackage{algpseudocode}
\usepackage{algcompatible}
\usepackage{enumerate}
\usepackage{hyperref}
\usepackage{cancel}
\usepackage{bbm}
\usepackage{dsfont}
\usepackage{adjustbox}
\usepackage{dblfloatfix} 
\usepackage{array}

\def\b0{{\mathbf{0}}}

\newmuskip\pFqmuskip

\def\argmax{\mathop{\rm \arg\!\max}}

\newtheorem{remark}{Remark}

\title{Antenna Selection for \\Large-Scale MIMO Systems with Low-Resolution ADCs}

\name{Jinseok Choi$^\dagger$, Junmo Sung$^\dagger$, Brian L. Evans$^\dagger$, and Alan Gatherer$^*$ 
\thanks{J. Choi, J. Sung, and B. L. Evans were supported by gift funding from Huawei Technologies.} 
}
\address{$^\dagger$Wireless Networking and Communications Group, The University of Texas at Austin\\
Email: \{jinseokchoi89, junmo.sung\}@utexas.edu, bevans@ece.utexas.edu\\
$^*$Systems and Design for Wireless Communications, Huawei, Plano, Texas, USA\\
Email: alan.gatherer@huawei.com}
\begin{document}
%
\maketitle

\begin{abstract}
One way to reduce the power consumption in large-scale multiple-input multiple-output (MIMO) systems is to employ low-resolution analog-to-digital converters (ADCs). 
In this paper, we investigate antenna selection for large-scale MIMO receivers with low-resolution ADCs, thereby providing more flexibility in resolution and number of ADCs.
To incorporate quantization effects, we generalize an existing objective function for a greedy capacity-maximization antenna selection approach.
The derived objective function offers an opportunity to select an antenna with the best tradeoff between the additional channel gain and increase in quantization error. 
Using the generalized objective function, we propose an antenna selection algorithm based on a conventional antenna selection algorithm without an increase in overall complexity.
Simulation results show that the proposed algorithm outperforms the conventional algorithm in achievable capacity for the same number of antennas.
\end{abstract}
\vspace{-0.5em}
\section{Introduction}
\label{sec:intro}


Large-scale MIMO systems have drawn considerable attention as a promising technology for next generation cellular systems because they offer orders of magnitude improvement in spectral efficiency \cite{ngo2013energy, hoydis2013massive, larsson2014massive}.
Practical challenges such as hardware cost and power consumption, however, arise due to the large number of antennas \cite{lu2014overview}.
Since the power consumption of ADCs, $P_{\rm ADC}$, scales exponentially in the number of quantization bits $b$, i.e., $P_{\rm ADC} \propto 2^b$ \cite{walden1999analog}, high-speed and high-resolution ADCs would be the primary power consumers.
Consequently, receivers with low-resolution ADCs (1-3 bits) have emerged as a possible solution to this problem \cite{mezghani2007ultra, mo2015capacity}.






Most prior work in low-resolution ADCs has focused on the case where the number of ADCs is same as the number of antennas \cite{mezghani2007ultra, mo2015capacity, rusu2015low, choi2016near} without considering the tradeoff between the number of quantization bits and the number of ADCs.
Accordingly, a reduced number of low-resolution ADCs has recently been investigated by using analog processing with phase shifters \cite{choi2017adc, mo2017hybrid, choi2017resolution}.
Analog processing with phase shifters, however, is mostly effective in a poor scattering channel environment with respect to reducing the number of ADCs \cite{bogale2014beamforming, el2014spatially}.
Moreover, phase shifters require additional hardware cost and power consumption \cite{mendez2016hybrid}.

For large-scale MIMO receivers, 
analog processing with phase shifters was compared against analog processing with switches, i.e., antenna selection, to reduce the number of ADCs \cite{mendez2016hybrid}.
It was shown in \cite{mendez2016hybrid} that antenna selection provides similar spectral efficiencies as analog processing with phase shifters for equal power consumption in millimeter wave (mmWave) channels.
The antenna selection used in \cite{mendez2016hybrid} does not exploit the sparsity of mmWave channels, whereas the analog processing with phase shifters does.
Hence, antenna selection is less limited by the channel environment such as low scattering than analog processing with phase shifters when reducing the number of ADCs. 
On that account, antenna selection is more applicable to general channels for the massive MIMO receiver. 
Indeed, for channels measured at $2.6$ GHz, a great number of ADCs could be turned off by using antenna selection without a substantial performance loss \cite{gao2015massive}.
Previously proposed antenna selection methods \cite{gao2015massive, gharavi2004fast, gorokhov2003receive, sadek2007active}, however, focused on MIMO systems without any quantization errors. 
Consequently, for low-resolution ADC receivers, a new antenna selection method that incorporates coarse quantization effect needs to be developed.

In this paper, we investigate antenna selection for large-scale MIMO systems with low-resolution ADCs.
We derive the tradeoff between the channel gain from selecting an antenna and its impact on quantization error.
In particular, we derive an antenna selection objective function for greedy-based antenna selection, which is a generalized version of the function in \cite{gharavi2004fast}.
The objective function measures $(i)$ the effect of quantization error from previously chosen antennas in the capacity gain that comes from selecting an additional antenna, and $(ii)$ the increase in quantization error from the additional antenna which acts as a penalty.
Accordingly, the tradeoff between the additional channel gain and the increase of the quantization error in antenna selection can be captured by using the derived objective function.
Leveraging the objective function, we propose a capacity-maximization antenna selection algorithm based on the conventional antenna selection algorithm without increasing the overall complexity.
Simulation results demonstrate that the proposed algorithm outperforms the conventional algorithm.


\section{System Model}
\label{sec:sys_model}


We consider a multi-user MIMO uplink system in which the BS with $N_r$ antennas serves $N_u$ users with single antenna.
We assume that the number of the BS antennas is much larger than the number of users, $N_r \gg N_u$.
Once the BS receives user signals, it selects $K$ antennas to use.
The selected antennas are connected to RF chains with ADCs.
Assuming a narrowband channel, the received baseband analog signal ${\bf r} \in \mathbb{C}^{N_r}$ can be expressed as
\begin{align}
	\label{eq:r}
	{\bf r} = \sqrt{\rho}{\bf Hs} + {\bf n}
\end{align} 
where $\rho$, $\bf H$, ${\bf s}$, and ${\bf n}$ denotes the transmit power, the $N_r \times N_u$ channel matrix, the user symbol vector, and the additive white Gaussian noise (AWGN) vector, respectively. 
We assume a zero mean and unit variance for $\bf s$ and ${\bf n}$. 
The $i$th column of ${\bf H}$ corresponds to the channel vector for user $i$, given as
\begin{align}
	\nonumber
	{\bf h}_i = \sqrt{\gamma_i}{\bf g}_i. 
\end{align}
The large scale fading gain $\gamma_i$ includes the geometric attenuation and shadow fading, and ${\bf g}_i$ denotes the vector of small scale fading gains for user $i$.
We use ${\bf f}^H_i$ to denote the $i$th row of ${\bf H}$.
We consider that the channel $\bf H$ is known at the BS and unknown at users.
Since the BS selects $K$ antennas and connects the selected antennas to $K$ RF chains, after the antenna selection, the received analog signal \eqref{eq:r} becomes
\begin{align}
    {\bf r}_\mathcal{K} = \sqrt{\rho} {\bf H}_\mathcal{K}{\bf s}+{\bf n}_\mathcal{K}
\end{align}
where ${\mathcal{K}}$ represents the index set of selected antennas with the cardinality of $|\mathcal{K}| = K$, ${\bf r}_\mathcal{K} \in \mathbb{C}^K$ denotes the received signal vector for the selected antennas, ${\bf H}_{\mathcal{K}} \in \mathbb{C}^{K\times N_u}$ is the channel matrix for the selected antennas in $\mathcal{K}$, 
and ${\bf n}_{\mathcal{K}} \in \mathbb{C}^K$ indicates the corresponding noise vector.

After the antenna selection, each real and imaginary component of the complex output $r_{\mathcal{K}(i)}$ is quantized at the pair of ADCs, where ${\mathcal{K}(i)}$ is the $i$th selected antenna.
For analytical tractability, the additive quantization noise model (AQNM) is used to linearize the quantization process as a function of quantization bits $b$.
AQNM \cite{orhan2015low} shows reasonable accuracy for b = 1, 2, and 3 in low and medium SNR ranges.
Adopting the AQNM, the quantized signal becomes
\begin{align} 
	\nonumber
	{\bf y} & = \mathcal{Q}\bigl({\rm Re}\{{\bf r}_\mathcal{K}\}\bigr)  + j\mathcal{Q}\bigl({\rm Im}\{{\bf r}_\mathcal{K}\}\bigr)\\ \label{eq:y}
	&= \alpha \sqrt{\rho} {\bf H}_\mathcal{K}{\bf s}+ \alpha {\bf n}_\mathcal{K} +{\bf q}
\end{align} 
where $\mathcal{Q}(\cdot)$ is the element-wise quantizer function.
Here, $\alpha$ is defined as $\alpha = 1- \beta $ and considered to be the quantization gain $(\alpha <1)$, and $\beta$ is the normalized mean squared quantization error $\beta = \frac{\mathbb{E}[|{y}_i - {y}_{{\rm q}i}|^2]}{\mathbb{E}[|{y}_i|^2]}$.
Assuming a scalar minimum mean squared error quantizer and Gaussian signaling for ${\bf s} \sim \mathcal{CN}(\mathbf{0}, \mathbf{I})$ where $\mathcal{CN}(\mathbf{0}, \mathbf{I})$ represents complex Gaussian distribution with a zero mean vector ${\bf 0}$ and the identity matrix ${\bf I}$ with proper dimension for a covariance matrix, $\beta$ is approximated as $\beta \approx \frac{\pi\sqrt{3}}{2} 2^{-2b}$ for $b > 5$.
Note that $b$ is the number of quantization bits for each real and imaginary part.
The values of $\beta$ for $b \leq 5$ are shown in Table \ref{tb:beta}.
The vector ${\bf q}$ represents the additive quantization noise and is uncorrelated with the quantization input ${\bf r}_\mathcal{K}$.
The quantization noise follows the complex Gaussian distribution with a zero mean ${\bf q} \sim \mathcal{CN}({\bf 0},{\bf R}_{\bf qq})$.
The covariance matrix $\mathbf{R}_{\bf qq}$ for the channel $ {\bf H}_{\mathcal{K}}$ is given by
\begin{align}
	\label{eq:cov}
	\mathbf{R}_{\bf qq}= \alpha(1-\alpha)\,{\rm diag}(\rho{\bf H}_\mathcal{K}{\bf H}_\mathcal{K}^H + {\mathbf{I}})
\end{align}
where ${\rm diag}(\rho{\bf H}_\mathcal{K}{\bf H}_\mathcal{K}^H + \mathbf{I})$ represents the diagonal matrix of the diagonal entries of $\rho{\bf H}_\mathcal{K}{\bf H}_\mathcal{K}^H + {\bf I}$.

\begin{table}[!t]
\centering
\caption{The Values of $\beta$ for Quantization Bits $b$ }
\label{tb:beta}
\begin{tabular}{ l c c c c c }
	\hline
	$ b$  & 1 & 2 & 3 & 4 & 5\\
  	\hline
 	$\beta$   & 0.3634 & 0.1175 & 0.03454 & 0.009497 & 0.002499 \\
 	\hline
\end{tabular}
\vspace{-0.5em}
\end{table}

\section{Antenna Selection}

\subsection{Performance Measure}
\label{sec:measure}

In this section, we examine the key difference of the antenna selection problem at the receiver with low-resolution ADCs from the conventional problem, and further propose an antenna selection method based on a capacity-maximization approach. 
Under the considered system in \eqref{eq:y}, the capacity for the given channel matrix ${\bf H}_\mathcal{K}$ can be expressed as
\begin{align}
	\label{eq:capacity}
	C({\bf H}_\mathcal{K}) = \log_2 \Big|{\bf I} + \rho\alpha^2\big( \alpha^2 {\bf I} + {\bf R}_{{\bf q}{\bf q}}\big)^{-1}{\bf H}_\mathcal{K}{\bf H}^H_\mathcal{K} \Big|
\end{align}
where ${\bf R}_{\bf qq}$ is given in \eqref{eq:cov}.
Note that in the system with low-resolution ADC, the quantization noise covariance matrix ${\bf R}_{\bf qq}$ is included in the capacity expression \eqref{eq:capacity}  as a penalty term for each antenna.
\begin{remark}
	\label{rm:intuition}
	Since each diagonal entry of ${\bf R}_{\bf qq}$ contains an aggregated channel gains at each selected antenna $\|{\bf f}_{\mathcal{K}(i)}\|^2$, the tradeoff between the channel gain and its influence on quantization error needs to be considered in antenna selection.
\end{remark}

\subsection{Capacity-Maximization Problem}
\label{sec:problem}

Using the capacity expression in \eqref{eq:capacity}, we formulate the antenna selection problem as follows:
\begin{align}
	\label{eq:problem}
	\mathcal{K}^\star = \argmax_{\mathcal{K} \subset \{1,\dots,N_r\}:|\mathcal{K}| = K}  C\big ({\bf H}_{\mathcal{K}}\big).
\end{align}
The large number of antennas $N_r$ makes it almost infeasible to search over all possible $\mathcal{K}$. 
Accordingly, to avoid an exhaustive search, we propose a quantization-aware antenna selection algorithm based on the greedy capacity-maximization approach \cite{gharavi2004fast}. 
The capacity in \eqref{eq:capacity} can be rewritten as 
\begin{align}
    &C({\bf H}_{\mathcal{K}})
	\label{eq:capacity2}
	 =  \log_2 \Big|{\bf I} +  \rho\alpha{\bf D}_{\mathcal{K}}^{-1}{\bf H}_{\mathcal{K}}{\bf H}^H_{\mathcal{K}} \Big|
\end{align}
where ${\bf D}_{\mathcal{K}} = {\rm diag}(1+\rho(1-\alpha)\|{\bf f}_{\mathcal{K}(i)}\|^2)$ is the diagonal matrix with $1+\rho(1-\alpha)\|{\bf f}_{\mathcal{K}(i)}\|^2$ for $i = 1, \dots, K$ at its diagonal entries.

Using \eqref{eq:capacity2}, we can express the capacity at the $(n+1)$th selection stage as follows:
\begin{align}
		\label{eq:capacity3}
    & C({\bf H}_{n+1}) \\ \nonumber
	& = \log_2 \Big|{\bf I} + \rho\alpha {\bf D}^{-1}_{n+1}{\bf H}_{n+1}{\bf H}^H_{n+1} \Big|\ \\ \nonumber
	& = \log_2 \Big|{\bf I} + \rho\alpha{\bf H}^H_{n+1} {\bf D}^{-1}_{n+1}{\bf H}_{n+1} \Big|\\ \nonumber
	& = \log_2 \biggl|{\bf I} + \rho\alpha\Bigl({\bf H}^H_{n} {\bf D}^{-1}_{n}{\bf H}_{n}\! +\! \frac{1}{d_{\mathcal{K}(n+1)}}{\bf f}_{\mathcal{K}(n+1)}{\bf f}^H_{\mathcal{K}(n+1)}\Bigr)\biggr|
\end{align}
\normalsize
where ${\bf H}_{n+1}$ is the channel matrix that corresponds to the $(n+1)$ selected antennas after the $(n+1)$th selection stage. 
Recall that ${\bf f}^H_{\mathcal{K}(n+1)}$ denotes the $(n+1)$th selected row of ${\bf H}$, and $d_{\mathcal{K}(n+1)}$ is the corresponding diagonal entry of ${\bf D}_{n+1}$.

We can decompose \eqref{eq:capacity3} into the capacity after the $n$th selection and the capacity increase that comes from selecting the $(n+1)$th antenna.
To this end, we use the matrix determinant lemma $|{\bf A} + {\bf u}{\bf v}^H| = |{\bf A}|(1+{\bf v}^H{\bf A}^{-1}{\bf u})$ as follows:
\begin{align}
	\nonumber
	& C({\bf H}_{n+1}) 
	= C({\bf H}_{n}) + \log_2\biggl(1\!+\!\frac{\rho \alpha}{d_{\mathcal{K}(n+1)}}c_{\mathcal{K}(n+1),n} \biggr)
\end{align}
where 
\begin{align}
	\label{eq:capacity gain}
	c_{\mathcal{K}(n+1),n} ={\bf f}^H_{\mathcal{K}(n+1)} \Bigl({\bf I}\! +\! \rho\alpha{\bf H}^H_{n} {\bf D}^{-1}_{n}{\bf H}_{n}\Bigr)^{-1}{\bf f}_{\mathcal{K}(n+1)}.
\end{align}
To maximize $C({\bf H}_{n+1})$ given the $n$ selected antennas, we need to find the next antenna $j$ which maximizes ${c_{j,n}}/{d_j}$. 
Accordingly, the antenna selection problem becomes
\begin{align}
	\label{eq:problem2}
    J =\argmax_{j }  \frac{c_{j,n}}{d_j}. 
\end{align}
Unlike the objective function of the capacity-maximization algorithm with no quantization error in \cite{gharavi2004fast}, the derived objective function $c_{j,n}/d_j$ incorporates $(i)$ the effect of the quantization error from the previously selected $n$ antennas to the next antenna $j$ in $c_{j,n}$, and $(ii)$ the additional quantization error from selecting the antenna $j$ as a penalty for the antenna $j$ in the form of $1/d_j$.
Since $d_j$ is a function of the aggregated channel gain for the $j$th antenna, $d_j = 1+\rho(1-\alpha)\|{\bf f}_j\|^2$, solving the problem \eqref{eq:problem2} gives the antenna $J$ which offers the best tradeoff between the channel gain from selecting an antenna and its influence on the increase of the quantization error.
This corresponds to the intuition in Remark \ref{rm:intuition}.
Note that \eqref{eq:problem2} is the generalized antenna selection objective function of the one in \cite{gharavi2004fast}; as the number of quantization bits $b$ increases, the quantization gain $\alpha$ increases as $\alpha \to 1$, which leads to $d_j \to 1$ and ${\bf D}_n \to {\bf I}$.

\begin{algorithm}[!t]
\caption{Quantization-Aware Fast Antenna Selection}
\label{algo:QAFAS}
\begin{enumerate}[1)]
	\item Initialize: $\mathcal{T} = \{1,\dots,N_r\}$ and ${\bf Q} = {\bf I}$.
	\item Initialize antenna gain and compute penalty: \\
		$c_j = \|{\bf f}_j\|^2$ and $d_j = 1 + \rho(1-\alpha)\|{\bf f}_j\|^2$ for $j \in \mathcal{T}$.
	\item Select antenna $J$ using \eqref{eq:problem2}: $J=  \argmax_{j\in \mathcal{T}}  c_j/d_j$.
	\item Update candidate set: $\mathcal{T} = \mathcal{T}\setminus\{J\}$.
	\item Compute: ${\bf a} = \bigl({c_{J} + \frac{d_J}{\rho \alpha}}\bigr)^{-\frac{1}{2}} {\bf Q} {\bf f}_J$ and ${\bf Q} = {\bf Q} - {\bf a}{\bf a}^H$.
	\item Update $c_j = c_j - |{\bf f}^H_j{\bf a}|^2$ for $j \in \mathcal{T}$.
	\item Go to step 3 and repeat until select $K$ antennas.
\end{enumerate}
\end{algorithm}

To propose a quantization-aware antenna selection algorithm, we modify the fast antenna selection algorithm in \cite{gharavi2004fast} by using the derived objective function in \eqref{eq:problem2} without increasing the overall complexity.
Unlike the algorithm for perfect quantization, the quantization error term $d_j$ needs to be computed in advance.
Then, at the selection step, the proposed algorithm adopts \eqref{eq:problem2} to incorporate the influence of the quantization error of the previously selected antennas and the candidate antenna.
To compute $c_{j,n}$ in \eqref{eq:capacity gain}, we define 
\begin{align}
	{\bf Q}_n =  \Bigl({\bf I}\! +\! \rho\alpha{\bf H}^H_{n} {\bf D}^{-1}_{n}{\bf H}_{n}\Bigr)^{-1}.
\end{align}
Then, ${\bf Q}_n$ can be updated efficiently by using the matrix inversion lemma as ${\bf Q}_{n+1} = {\bf Q}_n - {\bf a}{\bf a}^H$,
where ${\bf a} = \bigl({c_{J} + \frac{d_J}{\rho \alpha}}\bigr)^{-{1}/{2}} {\bf Q} {\bf f}_J
$.
Finally, $c_{j,n}$ can be updated as
\begin{align}
	\nonumber
	c_{j,n+1} &= {\bf f}^H_{j}{\bf Q}_{n+1}{\bf f}_{j} = {\bf f}^H_{j}\bigl({\bf Q}_n - {\bf a}{\bf a}\bigr)^H{\bf f}_{j}  
	\\ \nonumber
	&= c_{j,n} - |{\bf f}^H_j{\bf a}|^2.
\end{align}
The proposed algorithm, we call quantization-aware fast antenna selection (QAFAS), is described in Algorithm \ref{algo:QAFAS}.
Note that the complexity for step 5 and 6 are $O(KN_u^2)$ and $O(KN_uN_r)$, respectively.
Since we assume the large antenna arrays at the BS $(N_r \gg N_u)$, the overall complexity becomes $ O(K N_u N_r)$.
Thus, the proposed algorithm does not increase the overall complexity from the conventional algorithm for the perfect quantization in \cite{gharavi2004fast}, which provides the opportunity to be practically implemented without increase of computational burden.
In the following section, we demonstrate the performance of the proposed algorithm for the large-scale MIMO systems with coarse quantization.

%
%

\section{Simulation Results}

In this section, we evaluate the proposed algorithm and compare the algorithm with the fast antenna selection algorithm in \cite{gharavi2004fast} which shows a comparable performance to the optimal selection case under a perfect quantization assumption. 
The Matlab code is available at \cite{choi2017antennasoftware}.
We also include a random selection case for a reference in capacity performance.
We assume Rayleigh channel with a zero mean and unit variance.
For the large scale fading, we adopt the log-distance pathloss model \cite{erceg1999empirically}.
We consider randomly distributed users over a single cell with a cell radius of 2 $km$.
We assume the minimum distance between the BS and users to be $100$ $m$.
Considering a $2.4$ GHz carrier frequency with $10$ MHz bandwidth, we use $8.7$ dB lognormal shadowing variance and $5$ dB noise figure with $N_r = 128$ BS antennas and $N_u = 10$ users.

\begin{figure}[!t]\centering
	\includegraphics[scale = 0.32]{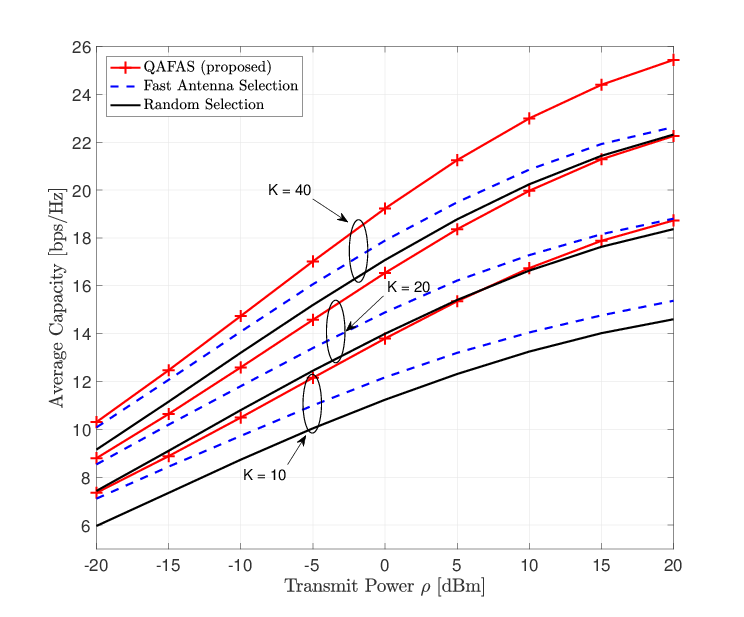}
	\caption{Average capacity for different transmit power with $N_r =128$ BS antennas, $N_u = 10$ users, $b = 3$ quantization bits, and $K \in \{10, 20, 40\}$ selected antennas.} 
	\label{fig:capacity power}
\end{figure}

Fig. \ref{fig:capacity power} shows the average capacity versus the transmit power $\rho$ for $b = 3$ quantization bits in $K\in \{10,20,40\}$ antenna selection cases.
The proposed QAFAS algorithm outperforms the conventional fast antenna selection method.
The gap between the algorithms increases as the transmit power $\rho$ increases because the quantization error becomes more dominant than the AWGN as the transmit power increases.
Accordingly, the proposed QAFAS algorithm which incorporates the quantization error when selecting antennas improves the capacity performance more when the quantization error becomes prevailing. 
Note that the QAFAS algorithm with $K \in \{10, 20\}$ achieves the capacity comparable to the random selection case with $K \in \{20, 40\}$ whose number of selected antennas $K$ is twice as many, respectively.

\begin{figure}[!t]\centering
	\includegraphics[scale = 0.32]{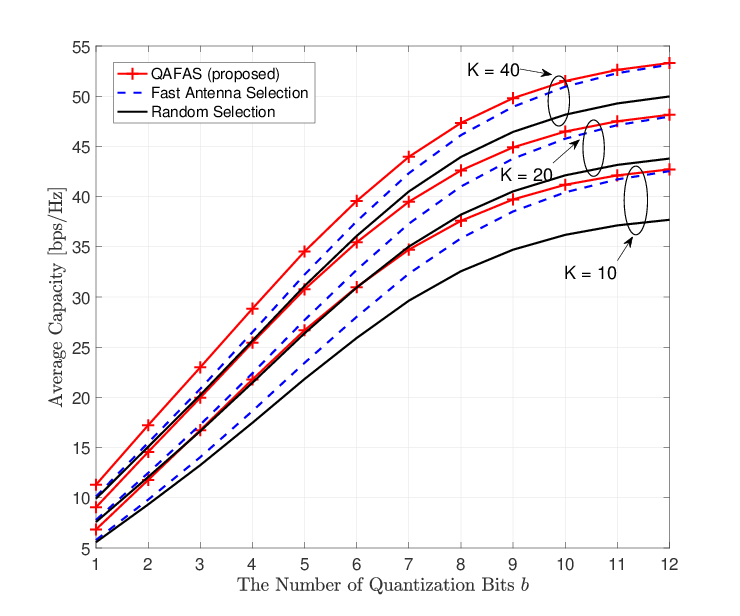}
	\caption{Average capacity for the different number of quantization bits with $N_r =128$ BS antennas, $N_u = 10$ users, $\rho = 5$ dBm transmit power, and $K \in \{10, 20\}$ selected antennas.} 
	\label{fig:capacity bit}
\end{figure}

For the different number of quantization bits $b$, the average capacities for $\rho = 5$ dBm transmit power in $K \in \{10, 20\}$ antenna selection cases are shown in Fig. \ref{fig:capacity bit}.
The proposed algorithm provides the highest capacity and improves the capacity compared to the random selection case in the low-resolution regime, whereas the conventional fast antenna selection marginally improves the capacity.
Notably, the capacity of the QAFAS algorithm shows relatively larger improvement compared to that of the fast antenna selection algorithm in the low-resolution regime than in the high-resolution regime, which corresponds to the motivation of this work.
In the low-resolution regime, the QAFAS algorithm with $K \in \{10, 20\}$ shows the comparable capacity to that of random selection with $K \in \{20, 40\}$, respectively.

\section{Conclusion}

In this paper, we proposed a new antenna selection algorithm for large-scale MIMO systems with low-resolution ADCs.
For each unselected antenna, the derived objective function measures the increase in capacity, like a conventional approach. 
Unlike a conventional approach, the proposed objective function measures the effect on capacity due to quantization error from previously chosen antennas and the increase in quantization error if an unselected antenna were chosen.
Since the derived objective function captures the tradeoff between the additional channel gain and the increase in quantization error for antenna selection, we used the objective function to propose a capacity-maximization antenna selection algorithm.
The simulation results validated the proposed antenna selection algorithm for low-resolution ADCs.
Therefore, using the proposed algorithm, the antenna selection for large-scale MIMO systems with low-resolution ADCs offers more flexibility in the resolution and number of ADCs with higher rates than other selection methods.

\clearpage
\bibliographystyle{IEEEtran}
\bibliography{icassp18.bib}
\end{document}